\documentclass[journal,dvipsnames]{IEEEtran}
\usepackage{amsmath,amsfonts}
\usepackage{algorithmic}
\usepackage{algorithm}
\usepackage{array}
\usepackage[caption=false,font=normalsize,labelfont=sf,textfont=sf]{subfig}
\usepackage{textcomp}
\usepackage{stfloats}
\usepackage{url}
\usepackage{verbatim}
\usepackage{graphicx}
\usepackage{cite}
\usepackage{microtype}
\usepackage{booktabs}
\usepackage{multirow}
\usepackage{adjustbox}
\usepackage{amssymb}
\usepackage{tikz}
\usepackage{pgfplots}
\pgfplotsset{compat=1.16}
\usetikzlibrary{pgfplots.groupplots,shapes,arrows,chains,fit,backgrounds}

\pgfplotsset{every axis/.append style={
        scaled y ticks = false,
        scaled x ticks = false,
        y tick label style={/pgf/number format/fixed}
    }
}
\usepackage{environ}
\makeatletter
\newsavebox{\measure@tikzpicture}
\NewEnviron{scaletikzpicturetowidth}[1]{%
  \def\tikz@width{#1}%
  \begin{lrbox}{\measure@tikzpicture}%
  \BODY
  \end{lrbox}%
  \pgfmathparse{#1/\wd\measure@tikzpicture}%
  \BODY
}
\makeatother

\newcommand{\ie}{i.e.\ }
\newcommand{\eg}{e.g.\ }

\begin{document}
\bstctlcite{IEEEexample:BSTcontrol}

\emergencystretch 2em

\title{Towards Probabilistic Dynamic Security Assessment and Enhancement of Large Power Systems}

\author{Frédéric~Sabot,
        Pierre-Etienne~Labeau,
        and~Pierre~Henneaux
\thanks{F. Sabot, P. E. Labeau, and P. Henneaux are with the Université libre de Bruxelles, 1000 Brussels, Belgium, e-mail: frederic.sabot@ulb.be.}
\thanks{This work has been prepared with the support of the Energy Transition Fund, project CYPRESS (https://cypress-project.be).}
}

\maketitle

\begin{abstract}
This paper proposes a novel methodology for probabilistic dynamic security assessment and enhancement of power systems that considers load and generation variability, N-2 contingencies, and uncertain cascade propagation caused by uncertain protection system behaviour. In this methodology, a database of likely operating conditions is generated via weather data, a market model and a model of operators' preventive actions. System states are sampled from this database and contingencies are applied to them to perform the security assessment. Rigorous statistical indicators are proposed to decide how many biased and unbiased samples to simulate to reach a target accuracy on the statistical error on the estimated risk from individual contingencies. Optionally, a screening of contingencies can be performed to limit the computational burden of the analysis. Finally, interpretable machine learning techniques are used to identify the root causes of the risk from critical contingencies, to ease the interpretation of the results, and to help with security enhancement. The method is demonstrated on the 73-bus reliability test system, and the scalability to large power systems (with thousands of buses) is also discussed.

\end{abstract}


\section{Introduction}
\IEEEPARstart{T}{he} security of a power system can be defined as its ability to withstand disturbances arising from faults and unscheduled removal of equipment without disturbing its customers~\cite{ReliabilityDefinition}. In a planning horizon, security assessment has traditionally been performed on a limited set of ``umbrella states" (\eg peak load without renewables, low load, etc.)~\cite{CIGREreviewOfTools}. However, with the increasing penetration of intermittent energy sources, it is becoming difficult to define a set of states that cover the main weaknesses of the system and that are at the same time reasonably likely to occur.

Moreover, current security assessment methodologies are still strongly based on the so-called ``N-1 security criterion" which specifies that power systems must be able to withstand the loss of a single element (among the N elements initially active) while continuing to supply consumers and keep operating conditions acceptable. However, higher order contingencies (also called N-k contingencies), while rarer than N-1 contingencies, have caused a large share of past unreliability events, including many blackouts~\cite{CascadingMethodoAndChallenges}.

The perceived need to consider many operating points and N-k contingencies in security assessments has led to a growing interest in complementing traditional \emph{deterministic} methodologies with \emph{probabilistic} ones. This is shown for example by the new regulation requiring European Transmission System Operators (TSOs) to develop a probabilistic approach for security assessment of power systems by 2027~\cite{ACER}, and the following data collection campaigns for probabilistic risk assessment launched by said TSOs~\cite{ENTSOE-PSA}.

Probabilistic methods consider many system states, each weighted by its probability of occurrence and assess the \emph{risk} (product of frequency and consequences) of potential disturbances. They thus allow to estimate the level of reliability of the system and to achieve a better trade-off between costs and reliability~\cite{GARPURdrivers_and_barriers}.

A challenge of probabilistic methods that does not exist with current deterministic methods is the difficulty to quantify the potential consequences of a disturbance. Indeed, with classical deterministic methods, consequences are categorised as either acceptable or unacceptable. So, in principle, a simulation can be stopped as soon as a load is disconnected or if voltage stays too low for an unacceptable amount of time, and the associated scenario be declared unacceptable. With probabilistic methods however, the simulation has to be run longer to determine if the system stabilises but in a slightly degraded state, or if a cascade occurs, and if a cascade occurs, when does it halt.

Cascading outages are notoriously challenging to simulate~\cite{CascadingMethodoAndChallenges} due to the many interacting cascading mechanisms to consider. A large part of the literature on cascading outages is based on the quasi-steady-state approximation and thus neglects short-term dynamics~\cite{BenchmarkingQSS}. However, in the past decades, there has been a growing share of blackouts that occurred in only a few minutes or even seconds which calls for time-domain simulations~\cite{cascadeAcceleration}. Generally speaking, stability issues are playing a growing role in modern power systems due to reduced system inertia caused by the introduction of inverter-based generation, pressures to operate the grid closer to its limits, increase of static limits through dynamic line rating and better conductors, etc.; so the importance of performing dynamic security assessment is also increasing.

Fast cascading outages are particularly difficult to model as many tripping events can occur in a short period of time, so small variations in the timing of protection operations (caused \eg by small measurement inaccuracies) can change the order in which protections operate, which actually operate, and thus strongly affect how the cascade propagates and its final consequences. The impact of uncertain protection behaviour has however only been considered in a few papers. In~\cite{PierreIEEEtran}, fast cascading outages are simulated with dynamic event trees where each possible realisation of a protection system behaviour leads to different possible branches. In~\cite{SequencesRelaySobol}, the order of protection system operation is encoded in a matrix that is then used with extended forms of variance-based sensitivity estimators to rank how sensitive cascading outages are to power system variables. While both approaches help in better understanding cascading outages and the effect of uncertainties, they are computationally expensive and thus cannot be applied to systematic analyses, \ie analysis were many credible contingencies are considered.

Computation time is indeed a major challenge for Probabilistic Dynamic Security Assessment (PDSA) of power systems as it has to consider significantly more system states (and potentially more disturbances) than a deterministic assessment. For example, in~\cite{EurostagDSA}, a PDSA was performed on the French power grid. The authors used a brute force approach, and therefore simulated 1980 credible contingencies in 9870 system states for a total of almost 20 million RMS simulations which took 23 hours to perform using 10,000 parallel cores. The analysis only considered N-1 contingencies (and 8 N-2 contingencies), so it would take even more computation time if a large set of N-2 contingencies was considered.

Brute-force sampling of operating conditions (or more precisely, sampling of operating conditions based on their historical or assumed probability density function) is known to be computationally expensive, and researchers have thus proposed more advanced techniques, such as importance sampling and directed walks to try and reduce the number of simulations required to obtain statistically accurate results~\cite{Bugaje, VorwerkWalk}. These techniques generally use a first batch of samples drawn from an unbiased distribution to estimate the location of the system security boundary (or of a so-called information-rich region around the security boundary), and latter batches are then biased towards this security boundary. The risk is that if parts of the security boundary is missed in the first batch, it will be even less likely to be found in the following batches. To mitigate this risk, this paper proposes rigorous statistical accuracy indicator to allow determining the optimal number of crude and biased samples to reach a target statistical accuracy.

The last key challenging aspect of PDSAs is that, since they require performing thousands to millions of simulations, interpreting the results of the assessment can be difficult as manual inspection of the results of all simulations is not possible. Consequently, it is also difficult to identify cost-effective measures to increase system security, creating a barrier between security assessment (quantification of system security and of main contributors to insecurity) and security management (optimal reduction of risk).

The aforementioned challenges are serious barriers to the application of PDSA methodologies to real grids, especially larger ones (with more than thousands of elements). This paper thus present a new probabilistic dynamic security assessment and enhancement methodology that alleviates those challenges. Our contributions are as follows:

\begin{itemize}
  \item We propose rigorous statistical accuracy indicators to determine the optimal number crude and biased samples of operating conditions to sample to reach a target accuracy on the estimated risk from individual contingencies (section~\ref{sec:sampling}).
  \item We use stability indicators to further reduce computation time with very limited impact on accuracy. In our test case, this allows for a reduction of computation time of a factor 2, although higher speed-ups could be obtained with better indicators (section~\ref{sec:screening}).
  \item We propose indicators to predict which scenarios lead to fast cascading outages that are very sensitive to the timing of protection system operations, and for which even small modelling inaccuracies or measurements errors in the protection systems can significantly impact the cascade evolution and its final consequences. This allows us to identify scenarios for which Monte Carlo are necessary to accurately estimate the scenario consequences, and to avoid them for the remaining scenarios (section~\ref{sec:protection}).
  \item From the results of the PDSA, we identify a small set of critical contingencies that contribute to a large share of the total risk. We then use simple interpretable machine learning (ML) techniques to identify the root causes that makes these contingencies critical, helping operators to efficiently mitigate the risk associated with these contingencies (section~\ref{sec:ML}).
  \item The applicability of the proposed methodology is demonstrated on a medium-scale power system, the 73-bus Reliability Test System (RTS), in a High-Performance Computing (HPC) environment, considering both N-1 and N-2 contingencies (section~\ref{sec:results}). The scalability to larger grids is also discussed (section~\ref{sec:scalability}).
\end{itemize}

The remainder of the paper is organised as follows. Section~\ref{sec:methodology} presents our proposed PDSA framework. Section~\ref{sec:testcase} and \ref{sec:results} respectively present the RTS test case and results. Section~\ref{sec:scalability} discusses the applicability of the proposed methodology to large grids, and section~\ref{sec:conclusion} concludes with a summary and perspectives. All the data and algorithms used in this work are available at \url{https://fredericsabot.github.io/Publications.html}.

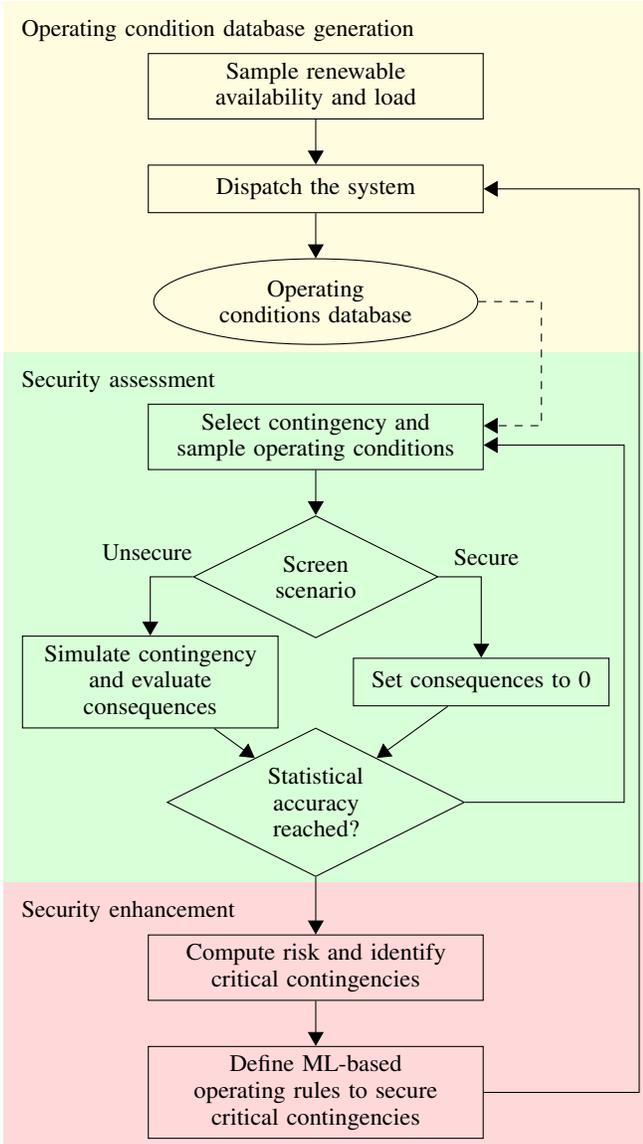
\begin{figure}
  \centering
  \begin{tikzpicture}[%
    >=triangle 60,              
    start chain=going below,    
    node distance=6mm and 20mm, 
    every join/.style={norm},   
    font=\small,
    ]
  \tikzset{
  base/.style={draw, on chain, on grid, align=center, minimum height=4ex},
  proc/.style={base, rectangle, text width=12em},
  shortproc/.style={base, rectangle, text width=9em},
  test/.style={base, diamond, aspect=2, text width=4em},
  term/.style={proc, rounded corners},
  storage/.style={base, ellipse, text width=8em},
  coord/.style={coordinate, on chain, on grid, node distance=6mm and 25mm},
  nmark/.style={draw, cyan, circle, font={\sffamily\bfseries}},
  norm/.style={->, draw},
  }
  \node [proc] (market) {Sample renewable availability and load};
  \node [proc, join]  (dispatch)    {Dispatch the system};
  \node [storage, join] (database) {Operating conditions database};

  \node [proc, below=18mm of database] (conditions) {Select contingency and sample operating conditions};
  \node [test, join] (screen) {Screen scenario};

  \node [shortproc, below left=14mm and 22mm of screen] (simulation) {Simulate contingency and evaluate consequences};
  \node [shortproc, below right=14mm and 22mm of screen] (safe) {Set consequences to 0};

  \node [test, below=30mm of screen]   (accuracy_reached)   {Statistical accuracy reached?};

  \node [proc, join, below=22mm of accuracy_reached]   (critical)   {Compute risk and identify critical contingencies};
  \node [proc, join]   (ML)   {Define ML-based operating rules to secure critical contingencies};

  \node [coord, left=of screen] (t1left)  {};
  \path (screen.west) to node [near start, yshift=1.5em, xshift=-1.5em] {Unsecure} (simulation);
  \draw [->] (screen.west) -| (simulation.north);
  \path (screen.east) to node [near start, yshift=1.5em, xshift=1.5em] {Secure} (safe);
  \draw [->] (screen.east) -| (safe);

  \draw [->] (simulation) -- (accuracy_reached);
  \draw [->] (safe) -- (accuracy_reached);

  \node [coord, right=41mm of accuracy_reached] (c2)  {};
  \draw [->] (accuracy_reached.east) -- (c2) |- ([yshift=-2mm] conditions);
  \node [coord, right=43mm of ML] (c5)  {};
  \draw [->] (ML.east) -- (c5) |- (dispatch);
  \node [coord, right=30mm of database] (c1)  {};
  \draw [->, dashed] (database.east) -- (c1) |- ([yshift=2mm] conditions);

  \node [above left=3mm and 18mm of market, anchor=west] (label1) {\parbox{6cm}{Operating condition database generation}};
  \node [above left=3mm and 18mm of conditions, anchor=west] (label2) {\parbox{6cm}{Security assessment}};
  \node [above left=3mm and 18mm of critical, anchor=west] (label3) {\parbox{6cm}{Security enhancement}};

  \coordinate[right=21mm of market] (aux1);
  \coordinate[right=21mm of conditions] (aux2);
  \coordinate[right=21mm of critical] (aux3);

  \begin{scope}[on background layer]
    \node[fill=yellow!15,fit=(label1) (aux1) (market) (dispatch) (database)]{};
    \node[fill=green!15,fit=(label2) (aux2) (conditions) (screen) (simulation) (safe) (accuracy_reached)]{};
    \node[fill=red!15,fit=(label3) (aux3) (critical) (ML)]{}; 
  \end{scope}

  \end{tikzpicture}
  \caption{Flowchart of the proposed PDSA methodology}
  \label{fig:flowchart}
\end{figure}

\section{Methodology}
\label{sec:methodology}

Our proposed methodology consists of three main steps as shown in the flowchart in Fig.~\ref{fig:flowchart}. The first step is the generation of a large database of likely system states for which security will be assessed. This step is a key element of any probabilistic analysis and is therefore well studied in the literature. This is discussed in section~\ref{sec:database}.

The second step is the security assessment. In this step, initial operating conditions are sampled from the database generated in the previous step and contingencies are applied to these initial states. Time-domain simulations are then used to determine if those contingencies are secure or if they can lead to cascading outages. In the latter case, time-domain simulations are also used to evaluate the potential consequences of these cascades.

An important question when doing MC simulations is how to efficiently sample and when to stop sampling. This is discussed in section~\ref{sec:sampling}.

To limit computation time, scenarios which are expected to be secure are screened out of the analysis. The stability indicators used for this purpose are described in section~\ref{sec:screening}.

As discussed above, fast cascading outages are difficult to model as they can be very sensitive to the timing of protection system operations. This is addressed in section~\ref{sec:protection}.

Finally, the third step consists in using the results of the security assessment to perform security enhancement (\ie to reduce the risk of unwanted load shedding). This is discussed in section~\ref{sec:ML}.

\subsection{Generating a database of credible system states}
\label{sec:database}

The main challenges in generating credible operating conditions of a power system is to adequately model the spatio-temporal correlations between renewable energy availability, load, and asset availability. Fortunately, this has been extensively studied in the literature. We therefore use a method strongly based on the one developed in the GARPUR project~\cite{StrathElia, StrathGARPUR}, and used by some European TSOs and ENTSO-E to perform adequacy studies~\cite{ACER_MC_year, EliaAdequacy}.

The methodology consists in, first, using weather data to generate so-called ``Monte Carlo (MC) years''. An MC year is time series realisation of renewable generation availability and load for one year with a typical resolution of one hour. Please refer to~\cite{StrathGARPUR} for more information on how to generate MC years while considering for asset outages and for the temporal and geographical correlations between renewable outputs and loads. Secondly, for each MC year, a market model is used to determine the commitment of thermal generators. Finally, each year is divided into (\eg hourly) snapshots, and a Security-Constrained Optimal Power Flow (SCOPF) is performed to guarantee that static limits are not exceeded in all possible N-1 conditions. All snapshots are saved in a database which will be sampled in the security assessment.

As this part of the workload is computationally inexpensive (compared to performing thousands of dynamic simulations), an arbitrarily large database of operating conditions can be generated.

Another approach to generate credible system states could be to try to estimate the multivariate probability density function of all stochastic variables based on historical measurements. This approach has been extensively studied during the iTesla project~\cite{KonstantelosCopulas, EurostagDSA}. In this project, the dispatch of the system (including potential operator actions and system topology) has also been inferred from historical data. The advantage of this method compared to the GARPUR approach is that it can potentially be more accurate since it is based on real historical data\footnote{The iTesla approach is especially effective at predicting the system topology (\ie substation configurations) for a given realisation of renewable availability and load, which is more difficult to model in an SCOPF.}. However, the reliance on historical data makes it a less flexible approach. In particular, it does not allow to model the impact of climate change on the likelihood of droughts and other severe weather events. Also, if the system or its operating rules are modified (\eg to enhance security as we will discuss in section~\ref{sec:ML}), historical data on operator actions might no longer be relevant. In this work, only the GARPUR approach was used.

\subsection{Sampling}
\label{sec:sampling}

Once the database of credible system states is generated, states are sampled and contingencies are applied to them in order to perform the security assessment. As for any MC algorithm, the main question is how to efficiently sample and how many samples to simulate to reach a target statistical accuracy with minimal computation burden\footnote{In the GARPUR approach, system states generated in the above steps are not sampled but clustered, and the analysis performed on the cluster centroids. The issue with this approach is that the number of clusters has to be defined before performing the analysis. Moreover, it is very difficult to quantify the error introduced by the clustering process, and thus to strike a good balance between the number of clusters and computation time.}. 

The standard MC approach is to sample system states proportionally to their likelihood until a stopping criterion is reached. Most papers stop sampling once they obtain satisfactory accuracy on the \emph{total} risk estimate. Here however, we argue that it is more useful to have an accurate estimation of the risk of \emph{individual} contingencies. Indeed, most security-enhancement actions (redispatches, system integrity protection schemes, synchronous condensers) solve local issues caused by a limited set of contingencies. To efficiently reduce the total risk, it is thus necessary to identify the most critical contingencies and to focus on reducing their individual contributions to the risk. In this work, a stopping criterion is thus defined for each contingency as

\begin{equation}
  \label{eq:stop}
  SE_i \leq \epsilon \; R
\end{equation}


\noindent where \(SE_i\) is the Standard Error (SE) of the risk of contingency \(i\), \(R\) is the total estimated risk, and \(\epsilon\) is a user-defined threshold (the choice of this threshold is discussed in more details in section~\ref{sec:results_sampling}). System states are thus sampled independently for each contingency until the SE of each contingency is smaller than a fraction of the total risk (\(\epsilon \, R\)). To be clear, all contingencies are enumerated (not sampled), and system states are sampled. This guarantees that the most critical contingencies can be correctly identified (if their contribution to the total risk is higher than \(\epsilon \, R\)).

The above criterion requires an estimate of the total risk. Thus, to warm up the algorithm, simulations are performed for all contingencies and for a few (\eg 5) operating conditions samples. This gives a first (rough) estimate of the total risk. Contingencies for which the stopping criterion (\ref{eq:stop}) is far from being satisfied are then sampled with higher priority. The estimate of the total risk is iteratively improved when more samples are simulated until (\ref{eq:stop}) is satisfied for all contingencies.

When using the standard MC approach, the SE can be computed via

\begin{equation}
  \label{eq:SE_classic}
  SE_i = f_i \sqrt{\frac{\sigma_i^2}{N_i}}
\end{equation}

\noindent where \(f_i\) is the frequency of contingency \(i\) and \(\sigma_i^2\) is the variance of its consequences. However, this variance is often unknown and therefore approximated by the sample variance \(\tilde{\sigma}_i^2\).
This approximation is commonly used but can be inaccurate especially for low probability high impact scenarios and for small values of \(N_i\). For example, if \(N_i\) samples of operating conditions are drawn for a given contingency and all show no consequences, the sampled variance will be zero, therefore the stopping criteria will be satisfied, and no other samples will be drawn. In this case, the confidence interval \([\tilde{\mu_i} - x SE_i, \tilde{\mu_i} + x SE_i]\) (where \(\tilde{\mu_i}\) are the sampled average consequences of contingency \(i\)) is infinitely narrow regardless of the value of \(x\) indicating perfect statistical accuracy. However, the \(N_i+1\)th sample might still lead to a blackout showing that the above approach might underestimate the risk.

To estimate the bias introduced in the above approach, it is useful to notice that if a contingency has a probability \(p_i\) to have consequences, the probability for \(N_i\) out of \(N_i\) samples to show no consequences is \((1-p_i)^{N_i}\). Therefore, if such samples are observed, then \(p_i\) satisfies

\begin{equation} \label{eq:proba}
  p_i < 1 - \sqrt[N_i]{1-\alpha} \ \text{with}\ \alpha \ \text{confidence}
\end{equation}

An upper bound on the bias is therefore \(f_i p_i M_C\) where \(M_C\) are the maximum consequences of a contingency (\eg a complete blackout).

In the more general case, the true mean \(\mu_i\) and variance \(\sigma_i\) of the consequences \(c_i\) of contingency \(i\) can be bounded by the sampled mean and variance of

\begin{equation}
  C_i = (1-p_i) \tilde{c_i} + p_i M_C
\end{equation}

\noindent where \(\tilde{c_i}\) is the sampled pdf of \(c_i\) and \(p_i M_C\) accounts for potential unsecure regions missed in the sampled \(\tilde{c_i}\). The following bounds can thus be obtained


\begin{IEEEeqnarray}{rCl}
  \mu_i & \leq & (1-p_i) \tilde{\mu_i} + p_i M_C \\
  \sigma_i^2 & \leq & (1-p_i) \tilde{\sigma_i^2} + p_i \beta_i^2 \ \text{where} \ \beta_i^2 = (M_C - \tilde{\mu_i})^2 \label{eq:sigma_bound}
\end{IEEEeqnarray}


For large \(N_i\) and \(\alpha = 95\%\), (\ref{eq:proba}) approximates to \(p_i < \frac{3}{N_i}\). Also, considering that \((1-p_i) \approx 1\) and injecting (\ref{eq:sigma_bound}) in (\ref{eq:SE_classic}), the following bound can be obtained for the SE of the risk of contingency \(i\)

\begin{equation}
  \label{eq:SE_bound}
  SE_i \leq f_i \sqrt{\frac{\tilde{\sigma}_i^2}{N_i} + \frac{3 \beta_i^2}{N_i^2}}
\end{equation}


This bound can then be used in the stopping criteria (\ref{eq:stop}). The first term in this bound is the classical variance term, and the second term represents how good is the ``coverage'' of the MC sampling, \ie how (un)likely it is to have missed important scenarios in the sampling process. The variance term accounts for the variance of the risk indicator, \ie how much the estimation will change if the computations are redone with a different random seed. And the coverage term is an upper bound on the bias of the estimator, \ie on average, how much the risk has been underestimated due to early termination of sampling process by the stopping criterion.

The demonstration above has been made for the case of a crude MC estimator (\ie sampling of system states based on their likelihood). Crude MC estimators can be slow to converge, especially in the presence of low frequency high impact scenarios, and more sophisticated methods have thus been developed. In the field of power systems, a commonly used method is importance sampling. It usually consists in biasing the sampling process towards unsecure cases in order to have more samples with consequences. In the context of resilience, this can be done for example by sampling more frequently the more severe earthquakes, because small earthquakes, while more frequent, will have a lower contribution to the risk as they often have low consequences.

In the context of security assessment, importance sampling is more difficult to use because it is difficult to know a priori if a given contingency will be less secure in the cases with high wind, the cases with high solar, and/or the cases with high/low load, etc. Adaptive importance sampling methods such as cross-entropy importance sampling circumvent this issue by drawing a first batch of samples in a crude MC way, identifying unsecure zones, then iteratively drawing additional batches of samples biased towards the unsecure zones. This approach can be dangerous because if an unsecure region is missed in the first batch, it will be increasingly more unlikely to be discovered it in the following batches.

Indeed, importance sampling (and other variance-reduction techniques) aim to reduce the variance of the MC estimator, but do not necessarily give better coverage. Actually, if the security region of a given contingency is not know a priori, crude MC is the most efficient sampling approach to minimise the risk associated with missed scenarios. Therefore, variance-reduction techniques can only be useful if

\begin{equation}
  \label{eq:coverage_requirement}
  \frac{\tilde{\sigma_i^2}}{N_i} > \frac{3 \beta_i^2}{N_i^2}
\end{equation}

However, as will be shown in section~\ref{sec:results} (notably in Fig.~\ref{fig:indic_N1} and \ref{fig:indic_N2}), the coverage term is dominant for most contingencies in our application, and crude MC is therefore the most effective approach.

Similarly, ML models could be trained using a first batch of MC simulations, and then used to speed up the simulation of the following MC samples. But again, if an unsecure zone is missed in the first batch, the ML model will not be able to predict it. So the criterion~(\ref{eq:coverage_requirement}) should also be satisfied before applying data-driven methods to speed up the MC simulation.


\subsection{Screening}
\label{sec:screening}

We just showed that crude MC sampling is the most efficient approach to guarantee adequate coverage of the sampling space for PDSA (at least for most contingencies). Because power systems are operated with a high level of reliability, a high share of MC simulations might be for secure and thus ``uninteresting'' scenarios. For example, the test system considered in this work is operated according to the N-1 criterion, yet more than 90\% of the N-2 scenarios (failure of 2 adjacent branches) do not lead to consequences.

This indicates that a significant speed-up can be obtained if secure scenarios are screened out of the analysis (up to a factor 10 in this case). Thus, in our PDSA framework, the security of each scenario (\ie operating conditions and contingency sample) is evaluated using a series of stability indicators. For unsecure scenarios, a time-domain simulation is performed to estimate the consequences of the scenario, while secure scenarios are simply skipped.

The stability indicators used in this work are as follows. For angle stability, the Critical Clearing Time (CCT) is estimated via the Extended Equal Area (EEA) method~\cite{EqualAreaCriterionEnergies} (using the critical cluster evaluation method from~\cite{EqualAreaCriterionPSCC}). A scenario is deemed unsecure is the actual clearing time is larger than the CCT plus a 50ms margin. The 50ms margin is used because it is preferable to have false positives (\ie secure scenarios that are predicted to be unsecure) rather than false negatives (\ie unsecure scenarios predicted secure). Indeed, for false positives, unnecessary simulations will be performed which increases computation time but not the risk estimate (because simulations of false positives will simply show that they are secure), while false negatives cause to underestimate the risk (because unsecure scenarios are disregarded).

The EEA method concerns the stability of synchronous generators. Some papers (\eg~\cite{ScreeningPLL}) argue that inverter-based generators can be modelled as synchronous machines with an inertia \(\frac{1}{K_i}\) where \(K_i\) is the gain of the integral component of the PLL. However, due to the large (\(>10\)s\textsuperscript{-1}) gains typically used, this leads to very low CCTs. However, this does not account for fault-ride through modes of inverters. Therefore, we modelled inverter-based generators as negative loads for the purpose of EEA.

For voltage stability, the indicator from~\cite{VoltageScreeningMachowski} is used, \ie a scenario is considered voltage secure if the short-circuit power at all buses is larger than 4 times the apparent power of the load of the bus. Regarding frequency stability, based on preliminary simulations on our test system, a scenario is deemed secure if it leads to a rate of change of frequency lower than 0.4~Hz/s and of loss of power generation less than 70\% of the primary reserve. Generators near a fault are allowed to disconnect if the fault lasts longer than 150ms.

\subsection{Handling of uncertain protection behaviour during fast cascading outages}
\label{sec:protection}

Once a scenario is sampled and passes the screening process, it is simulated to estimate its consequences. Some scenarios will lead to cascading outages which are difficult to accurately simulate~\cite{CascadingMethodoAndChallenges}. In particular, simulating fast cascading outages, \ie cascading outages lasting a few seconds to a few minutes, is challenging as many protection systems might operate in quick succession in a given cascade and the cascading path might thus be very sensitive to the timing of protection system operations (as small changes in the timing of protection operations can change the order of protection operations and which protections operate). One way to handle this complexity is to perform many MC simulations for each scenario with random parameters in all protection systems (with a small variance in the protection threshold to account for small measurement inaccuracies, and in the protection delays to account for the variance of circuit breaker opening time (variance of the order of a cicle)). However, this would further increase the already high computational burden of PDSAs.

\begin{figure}
  \centering
  \subfloat[]{%
      \begin{tikzpicture}
      \draw (0,0) -- (7,0);

      \foreach \x in {0,1,4.5,6,7}
        \draw (\x cm,3pt) -- (\x cm,-3pt);

      \draw (1,0) node[below=3pt] {$ 1_a $};
      \draw (4.5,0) node[below=3pt] {$ 2_a $};
      \draw (6,0) node[below=3pt] {$ 3_a $};
      \end{tikzpicture}
      \label{subfig:slowSeq}
  } \\ \vskip 0.5 \baselineskip
  \subfloat[]{%
      \begin{tikzpicture}
      \draw (0,0) -- (7,0);

      \foreach \x in {0,1,4,5.5,7}
        \draw (\x cm,3pt) -- (\x cm,-3pt);

      \draw (1,0) node[below=3pt] {$ 1_b $};
      \draw (4,0) node[below=3pt] {$ 2_b $};
      \draw (5.5,0) node[below=3pt] {$ 3_b $};
      \end{tikzpicture}
      \label{subfig:negative}
  } \\ \vskip 0.5 \baselineskip
  \subfloat[]{%
      \begin{tikzpicture}
      \draw (0,0) -- (7,0);

      \foreach \x in {0,1,2,3,7}
        \draw (\x cm,3pt) -- (\x cm,-3pt);

      \draw (1,0) node[below=3pt] {$ 1_c $};
      \draw (2,0) node[below=3pt] {$ 2_c $};
      \draw (3,0) node[below=3pt] {$ 3_c $};
      \end{tikzpicture}
      \label{subfig:order}
  } \\ \vskip 0.5\baselineskip
  \subfloat[]{%
      \begin{tikzpicture}
      \draw (0,0) -- (7,0);

      \foreach \x in {0,1,4.5,6,6.5,7}
        \draw (\x cm,3pt) -- (\x cm,-3pt);

      \draw (1,0) node[below=3pt] {$ 1_d $};
      \draw (4.5,0) node[below=3pt] {$ 2_d $};
      \draw (6,0) node[below=3pt] {$ 3_d $};
      \draw (6.5,0) node[below=3pt] {$ 4_d $};
      \end{tikzpicture}
      \label{subfig:new_event}
  }
  \label{fig:powertech_indicator}
  \caption{Prediction of the relevance of protection-related uncertainties using the indicator from~\cite{HandlingProtections}. (a) Slow sequence (reference). (b) Fast sequence for which the system is unlikely to be affected by protection-related uncertainties. (c) Fast sequence likely to be affected: \(3_c\) occurs before \(2_a\). (d) Fast sequence likely to be affected: a new event (\(4_d\)) occurs.}
\end{figure}
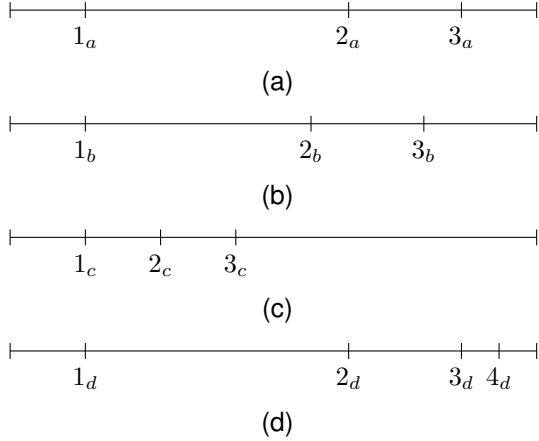

In previous work~\cite{HandlingProtections}, we proposed an indicator to predict if protection-related uncertainties can impact the cascading path for a given contingency and given operating conditions. The indicator is computed by simulating the system with two sets of protection systems. The first set is given values from the pdfs of the protection parameters that lead to the slowest possible operation of the protection systems. And the second set is given values that lead to the fastest possible operation. The second set is however not connected to circuit breakers to not affect the system evolution. For a given contingency and operating condition, one thus obtains two possible sequences of tripping events (one for each set of protection systems) from a single simulation. Protection-related uncertainties are expected to affect the consequences of a contingency if one event occurs in one sequence but not the other (Fig.~\ref{subfig:new_event}), or if comparing the two sequences shows the possibility for two events to be swapped (Fig.~\ref{subfig:order}). This indicator showed a very good accuracy in~\cite{HandlingProtections}. 

Thus, for each sampled scenario, we perform a first simulation to estimate if protection-related uncertainties can affect the cascading path. If they can, additional MC simulations are performed for this scenario with random protection parameters to estimate the most likely cascading paths and statistical indicators (average consequences, etc.). If they cannot, the consequences of the scenario are simply estimated from the initial simulation.

\subsection{Security enhancement}
\label{sec:ML}

Once the security assessment is performed, we have an estimate of the total risk of cascading outages and of the contribution of all contingencies to this risk. Therefore, critical contingencies can be identified and resources focused on those contingencies in order to efficiently reduce the risk. In our test case, 10 contingencies (out of 708) contribute to more than 40\% of the total risk. However, even when looking only at those 10 contingencies, hundreds to thousands of simulations might still have been performed to account for the variability of operating conditions, so it is not obvious to identify the main drivers of instability and how to best reduce the risk.

ML techniques, and in particular Decision Trees (DTs), have been used for many decades to identify the security boundary, \ie separation between secure and unsecure operating conditions, of a system based on the results of offline simulations~\cite{DT_Wehenkel}. This has often been done with the goal to define security rules and to help operational planners dispatch their grid in a more secure way. In this paper, we use the same techniques, but with the goal to identify the root causes of the risk from the critical contingencies. This allows long-term planning to better interpret the results of the PDSA and helps them to identify efficient security-enhancement actions (installation of system integrity protection schemes, synchronous condensers, etc.).

As in long-term planning, ``exact'' time-domain simulations are available, we decided to put more weight on the interpretability of the ML models used compared to their accuracy. In this case, we used linear Support Vector Machine (SVM) models to identify the security boundary of each individual critical contingencies. SVMs split the feature space by a hyperplane with most of the secure operating points on one side of the hyperplane and most unsecure points on the other side. The drawbacks of SVMs are that they are hard to visualise in high dimensional (\(>3\)) feature spaces and that the feature weights have no clear interpretation when there are strong correlations between features. To avoid those drawbacks, we used sequential feature selection to limit the dimension of the feature space. It consists in training an SVM for each feature individually, keeping the best one, adding a second feature, keeping the best one, etc. Section~\ref{sec:results_ML} demonstrates how this simple ML model \emph{and} the sequential feature selection procedures can identify the root causes of the risk from the critical contingencies.

It should be noted that the stopping criterion (\ref{eq:stop}) (and (\ref{eq:SE_bound})) of the PDSA guarantees that no important unsecure region has been missed in the sampling process. This is a necessary but not sufficient condition to train accurate models. In the PDSA, operating conditions are sampled based on their pdf. According to~\cite{Bugaje} however, sampling should be biased towards the security boundary and towards unlikely states for best training. For the sake of simplicity, we only used crude MC sampling in this work. However, as we focus on a limited set of critical contingencies (identified in the security assessment), the computation cost of simulating additional samples would be limited compared to the total computation cost of the PDSA. We therefore used a minimum of 1000 operating condition samples to train SVMs for each critical contingency.

\section{Test case}
\label{sec:testcase}

The test system used in this work is the Reliability Test System as defined by the Grid Modernization Lab Consortium (RTS-GMLC)~\cite{RTS-GMLC}. This version is similar to the RTS-96 but with a large part of the coal and nuclear fleet replaced with renewable generation and gas, making it more representative of modern grids. Additionally, the system has been mapped to a region in the southwestern US to define load and renewable output time series.

\begin{figure}
    \centering
    \includegraphics[width=\linewidth]{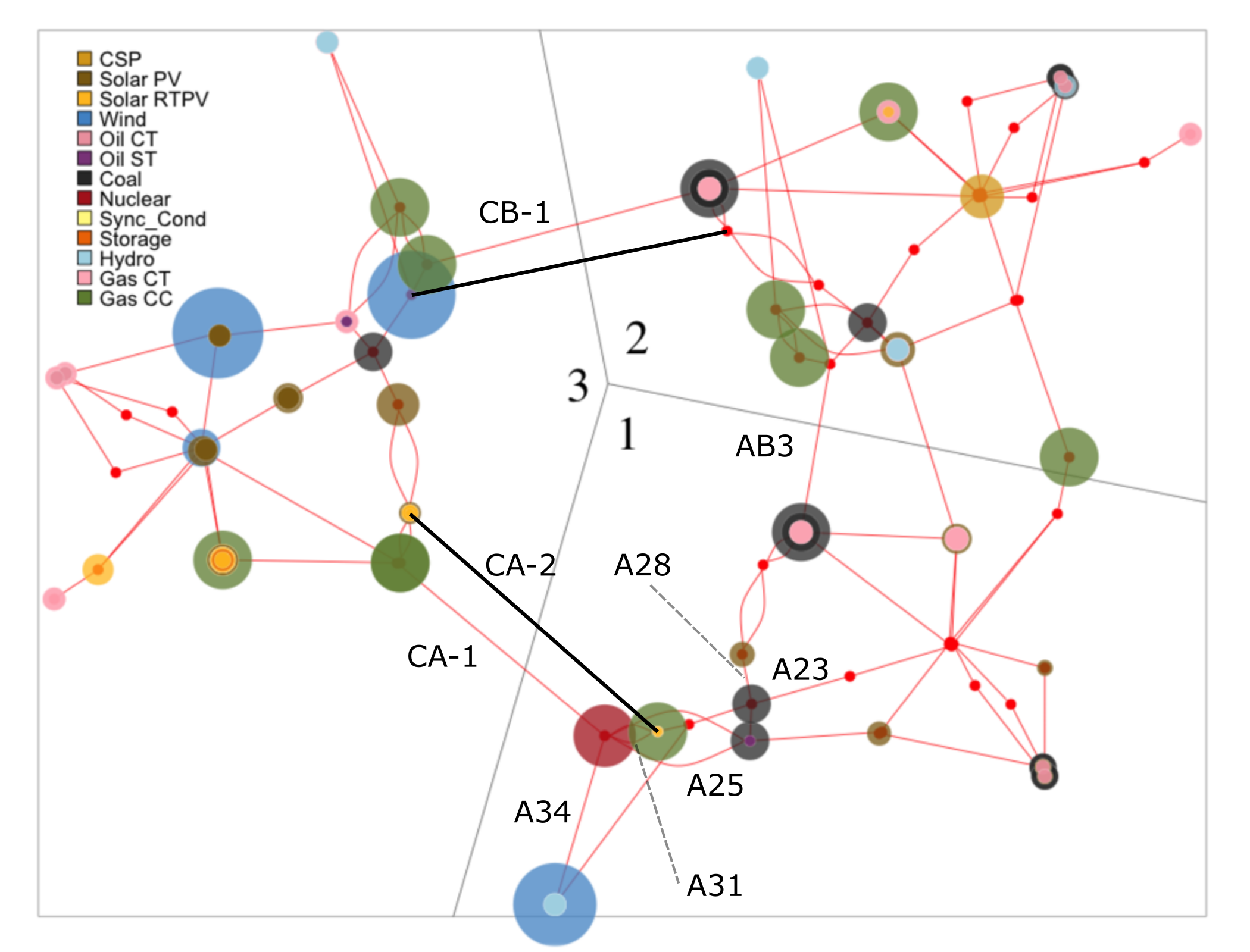}
    \caption{Network layout of the RTS-GMLC from~\cite{RTS-GMLC}. New interconnections are represented in black.}
    \label{fig:RTS}
\end{figure}

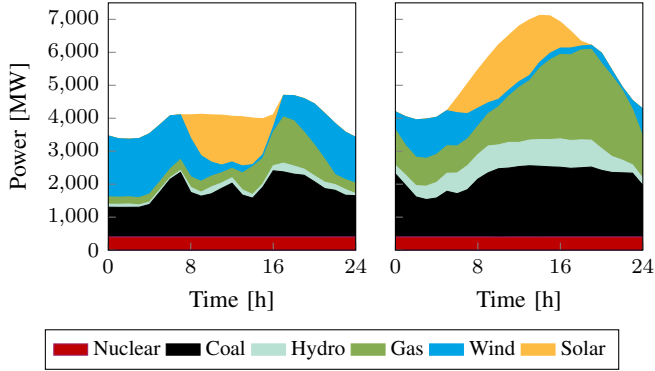
\begin{figure}
\centering
\begin{tikzpicture}
\pgfplotsset{
/pgfplots/area cycle list/.style={/pgfplots/cycle list={%
{RedViolet,fill=red!75!black,mark=none},%
{black,fill=black,mark=none},%
{SeaGreen!40!white,fill=SeaGreen!40!white,mark=none},%
{LimeGreen!80!black,fill=LimeGreen!80!black,mark=none},%
{Cerulean,fill=Cerulean,mark=none},%
{Dandelion,fill=Dandelion,mark=none},%
}
},
}
\begin{groupplot}[
    group style={%
        columns=2,
        group name=plots,
        xlabels at=edge bottom,
        y descriptions at=edge left,
        horizontal sep =15pt,
    },
    footnotesize,
    xlabel = {Time [h]},
    ylabel = {Power [MW]},
    stack plots=y,%
    area style,
    xmin=0, xmax=24,
    xtick={0,8,16,24},
    ymin = 0, ymax=7500,
    tickpos=left,
    width=0.55\linewidth,
    height=0.55\linewidth,
    every legend image post/.append style={
        scale=0.75
    }
]
\nextgroupplot
\addplot table [x=hour,y=nuclear] {dispatch_january.txt}
\closedcycle;
\addplot table [x=hour,y=coal] {dispatch_january.txt}
\closedcycle;
\addplot table [x=hour,y=hydro] {dispatch_january.txt}
\closedcycle;
\addplot table [x=hour,y=gas] {dispatch_january.txt}
\closedcycle;
\addplot table [x=hour,y=wind] {dispatch_january.txt}
\closedcycle;
\addplot table [x=hour,y=solar] {dispatch_january.txt}
\closedcycle;

\nextgroupplot[legend to name={DispatchLegend},legend style={legend columns=6}]
\addplot table [x=hour,y=nuclear] {dispatch_july.txt}
\closedcycle;
\addplot table [x=hour,y=coal] {dispatch_july.txt}
\closedcycle;
\addplot table [x=hour,y=hydro] {dispatch_july.txt}
\closedcycle;
\addplot table [x=hour,y=gas] {dispatch_july.txt}
\closedcycle;
\addplot table [x=hour,y=wind] {dispatch_july.txt}
\closedcycle;
\addplot table [x=hour,y=solar] {dispatch_july.txt}
\closedcycle;

\legend{Nuclear, Coal, Hydro, Gas, Wind, Solar}

\end{groupplot}

\node[below] at (current bounding box.south)
      {\pgfplotslegendfromname{DispatchLegend}};

\end{tikzpicture}
\caption{Market dispatch for a typical day in January (left) and July (right)}
\label{fig:dispatch}
\end{figure}

For this work, two interconnections have been added to limit curtailment of the (very) large wind plants in zone 3 as shown in Fig.~\ref{fig:RTS}. Market dispatches for typical days in January and July are shown in Fig.~\ref{fig:dispatch}. The market model used is a locational marginal pricing (LMP) market model developed in~\cite{Prescient}. Fig.~\ref{fig:dispatch} shows that high wind penetration (above 60\%) are sometimes reached especially for winter months during which load is relatively low.

As there was no dynamic data in the original RTS-GMLC, dynamic models of loads and generators have been added in this work: synchronous generators models are based on annex D of~\cite{vittalBook} and inverter-based generator models are based on~\cite{ChaspierreThesis}. The protection schemes modelled are the same as in~\cite{HandlingProtections}. They consist in distance protection of lines (with a load blinder), an under-frequency load shedding scheme, and under-voltage and loss-of-synchronism protection of generators. Since small variations in the timing of protection operations can lead to different cascading paths (with potentially different consequences), the uncertainty of protection behaviour is considered by associating probability density functions to their parameters as in~\cite{HandlingProtections}. Most notably, the opening time of circuit breakers can vary in [70, 90] ms, and the measurement of apparent impedances (for distance protections) is considered to have an accuracy of 10\%.

The contingencies considered in this work are three-phase faults occurring at one of the two extremities of lines. These faults are normally cleared by opening the faulted line in 100ms (N-1 contingency). However, we also consider that there is a 0.1 chance that the primary protection fails at that the fault is thus cleared in 200ms (N-1 contingency with delayed clearing). Also, we consider a 0.01 chance that one breaker fails to open when clearing the fault. We assume that breaker failure protection~\cite{HorowitzBook} is installed in all substations and that they are able to clear the fault by opening a single adjacent line in 200ms (leading to an N-2 contingency). It has been checked that the system is always secure for N-1 contingencies (with normal clearing time), these have thus not been considered in the PDSA. Only faults at the highest voltage level are considered which leads to a contingency list of 114 N-1 (with delayed clearing) and 594 N-2 contingencies. The frequency of line faults is taken as 2.5 faults per 100~km of line per year~\cite{FaultStatisticsFrance}\footnote{Due to lack of data, line fault probability is assumed independent of weather conditions. Fault and protection failure statistics significantly vary from country to country due to different weather and reliability and reporting  practices (\eg \cite{GridPSA} reports a 0.27 fault per year and per 100~km of 400~kV overhead lines for the Finnish grid, while \cite{FaultStatisticsFrance} reports 2.5 in France). Data collection plans (as initiated by ENTSO-E~\cite{ENTSOE-PSA}) and failure mode and effect analysis (as demonstrated in~\cite{GridPSA}) should thus be performed to be able to fully trust the results of a PDSA.}. Dynamic simulations are performed using Dynawo~\cite{Dynawo} on 10 32-cores AMD EPYC Rome 7542 CPU's at 2.9 GHz.

The contingencies considered in this work are line faults which are not cleared by the primary protection system due to protection failure or ``missed trips''. Another important class of protection failures are ``unwanted trips'', \ie trips that are not necessary to clear the fault and that thus unnecessarily disconnect elements. Unwanted trips that occur following a fault can also cause high-order contingencies and therefore have a significant contribution to the risk. However, the modelling of unwanted trips is more complex than for missing trips and will thus be studied in future work. Generally speaking, there is a large gap in the current literature regarding the modelling and probability estimation of contingencies. Indeed, many researchers perform security assessment studies considering N-2 contingencies caused by independent failures while this type of contingency is very rare, and the majority of historical blackouts has actually been caused by single contingencies that were exacerbated by hidden failures or other aggravating factors~\cite{CascadingMethodoAndChallenges}.

The simulation of a given scenario provides with an estimate of the consequences of a contingency in terms of MW of load shed. To translate this value in terms of societal cost, we used the simple restoration model and value of loss load from~\cite{VOLL}. For the RTS and at average load (4350~MW), the cost of a complete blackout is thus estimated at 500M€. This is the value used for \(M_C\).

\section{Results}
\label{sec:results}

This section presents the results of the application of our methodology to the RTS-GMLC system. This section is organised similarly to the methodology section with section~\ref{sec:results_sampling} discussing the sampling process and computational burden of the PDSA, section~\ref{sec:results_screening} discussing the performance of the screening process and its impact on accuracy and computation time, section~\ref{sec:results_protection} analysing the impact of protection-related uncertainties on fast cascading outages, and section~\ref{sec:results_ML} demonstrating how ML techniques can help understand the results of a PDSA.

\subsection{Sampling}
\label{sec:results_sampling}

A first PDSA has been performed without screening of scenarios to be used as a reference. The main results of this analysis are given in Table~\ref{tab:summary-N1N2}. It shows that (delayed clearing) N-1 contingencies and N-2 contingencies have a similar contribution to the total risk. Also, while there are more N-2 contingencies than N-1 contingencies (594 vs 114), N-2 contingencies require fewer simulations, and therefore less computation time than N-1 contingencies.

\begin{table}
\centering
\caption{Contribution to total risk and computation time of delayed clearing N-1 contingencies and N-2 contingencies (without screening)}
\label{tab:summary-N1N2}
\begin{adjustbox}{max width=\linewidth}
\begin{tabular}{@{}lll@{}}
\toprule
& N-1  & N-2 \\ \midrule
Risk (M€/y) & 8.6 & 12.4 \\
Number of simulations            & 97,503 & 59,648 \\
Computation time (core-h)        & 248  & 157 \\
Average computation time per simulation (s) & 9.1 & 9.5 \\ \bottomrule
\end{tabular}
\end{adjustbox}
\end{table}

This is because N-1 contingencies are contingencies that are more frequent but infrequently lead to significant consequences. It is thus necessary to sample many operating conditions to obtain a statistically accurate risk and guarantee a sufficient coverage of the likely operating conditions.

Fig.~\ref{fig:indic_N1} (resp. Fig.~\ref{fig:indic_N2}) shows the number of simulations performed for all N-1 (resp. N-2) contingencies and the associated standard error (decomposed in terms of variance and coverage). It shows that for most contingencies the coverage part of the SE is dominant compared to the variance part. For these contingencies, the SE bound reduces to

\begin{equation}
  SE_i \lesssim \frac{f_i}{N_i} \sqrt{3 \beta_i^2}
\end{equation}

\noindent and the number of simulations needed to satisfy the stopping criterion (\ref{eq:stop}) is thus directly proportional to the frequency of the contingency. For contingencies with non-negligible variance, the number of simulations needed is higher which explains the spikes of \(N_i\) in Fig.~\ref{fig:indic_N1} and \ref{fig:indic_N2}.

\begin{figure}
\centering
\begin{tikzpicture}
\pgfplotsset{width=0.9\linewidth,
        legend style={font=\footnotesize},}
\begin{groupplot}[
    group style={
        group name=my plots,
        group size=1 by 2,
        xlabels at=edge bottom,
        xticklabels at=edge bottom,
        vertical sep=20pt
    },
    footnotesize,
    width=8cm,
    height=4cm,
    xlabel=Contingency id,
    xmin=1, xmax=114,
    ymin=0,
    tickpos=left,
    ytick align=outside,
    xtick align=outside,
]
\nextgroupplot[ylabel=\(N_i\), height=3cm]
\addplot+ [mark=none, line width=1pt] table [x=id,y=N_static] {N_N1.txt};

\nextgroupplot[
    ylabel style={align=center}, ylabel=Statistical accuracy\\{[M€/y]},
    legend style={legend columns=2}, ymax=0.34]

\addplot+ [ybar interval, fill, mark=none] table [x=id,y expr=\thisrow{indic_1}*1.609344] {N_N1.txt};  
\addplot+ [mark=none] table [x=id,y expr=\thisrow{indic_2}*1.609344] {N_N1.txt};
\addlegendentry{Variance}
\addlegendentry{Coverage}
\end{groupplot}
\end{tikzpicture}
\caption{Number of sampled operating conditions and statistical accuracy for all delayed-clearing N-1 contingencies sorted in decreasing order of likelihood}
\label{fig:indic_N1}
\end{figure}
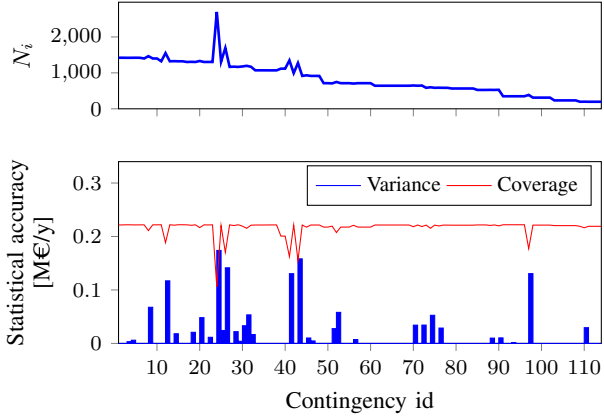
\begin{figure}
\centering
\begin{tikzpicture}
\pgfplotsset{width=0.9\linewidth,
        legend style={font=\footnotesize},}
\begin{groupplot}[
    group style={
        group name=my plots,
        group size=1 by 2,
        xlabels at=edge bottom,
        xticklabels at=edge bottom,
        vertical sep=20pt
    },
    footnotesize,
    width=8cm,
    height=4cm,
    xlabel=Contingency id,
    xmin=1, xmax=594,
    ymin=0,
    tickpos=left,
    ytick align=outside,
    xtick align=outside,
]
\nextgroupplot[ylabel=\(N_i\), height=3cm]
\addplot+ [mark=none, line width=1pt] table [x=id,y=N_static] {N_N2.txt};

\nextgroupplot[
  ylabel style={align=center}, ylabel=Statistical accuracy\\{[M€/y]},
  legend style={legend columns=2}, ymax=0.34]

\addplot+ [ybar interval, fill, mark=none] table [x=id,y expr=\thisrow{indic_1}*1.609344] {N_N2.txt};
\addplot+ [mark=none] table [x=id,y expr=\thisrow{indic_2}*1.609344] {N_N2.txt};

\addlegendentry{Variance}
\addlegendentry{Coverage}
\end{groupplot}
\end{tikzpicture}
\caption{Number of sampled operating conditions and statistical accuracy for all N-2 contingencies sorted in decreasing order of likelihood}
\label{fig:indic_N2}
\end{figure}
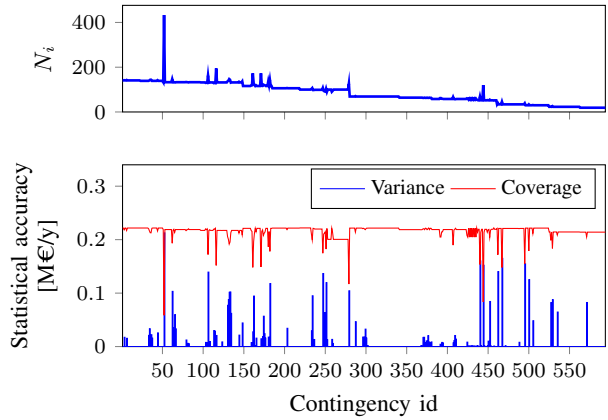

It is interesting to see that when aiming to minimise the SE of the risk contribution of individual contingencies, the best strategy is basically to use a crude MC approach (\ie sampling contingencies proportionally to their frequency of occurrence) (except for a few contingencies with high variance). In a crude MC approach, the total risk can be estimated as

\begin{equation}
  R = \left(\sum_i f_i\right) \left(\frac{1}{N} \sum_s c_s\right)
\end{equation}

\noindent where \(c_s\) are the consequences of the \(s\)th sample (random contingency and initial state). Using the same development as to derive Eq.~\ref{eq:SE_bound}, the following bound can be obtained for the SE of the total risk.

\begin{equation}
  \label{eq:SE_total}
  SE \leq \left(\sum_i f_i\right) \sqrt{\frac{\tilde{\sigma}^2}{N} + \frac{3 \beta^2}{N^2}}
\end{equation}

\noindent where \(\tilde{\sigma}\), \(\beta\), and \(N\) have the same definitions as \(\tilde{\sigma_i}\), \(\beta_i\), and \(N_i\) but for the total risk instead of individual contingencies. Fig.~\ref{fig:SE_total} shows how this SE evolves with the number of samples. After 150,000 samples (roughly the number of simulations performed in this study, cf. Table~\ref{tab:summary-N1N2}), the coverage term of SE is 4.3 smaller than the variance term while it was strongly dominant for the SE of individual contingencies. This is because the coverage term accounts for the likelihood of having missed unsecure regions during sampling, and while this likelihood is relatively high for individual contingencies, it is unlikely to miss unsecure regions for all of them. Fig.~\ref{fig:SE_total} shows the coverage term becomes smaller than the variance term after roughly 8000 samples. Thus, if one is only interested in the total risk and not in the risk of individual contingencies (for some reason), then variance-reduction techniques become viable after this point.

\begin{figure}
\centering
\begin{tikzpicture}
\pgfplotsset{width=0.9\linewidth,
        legend style={font=\footnotesize}}
\begin{axis}[
    xlabel={Number of samples},
    xmin=1000, xmax=150000,
    ymin=0, ymax=8,
    xmode=log,
    grid,
    log ticks with fixed point,
    ylabel= {SE [M€/y]},
    legend cell align=left,
    legend style={at={(1,1)},anchor=north east},
    ]

    \addplot+ [mark=none, dashed] table [x=x,y expr=\thisrow{sigma}*1.609344] {total_risk_SE.txt};
    \addplot+ [mark=none, dashdotted] table [x=x,y expr=\thisrow{coverage}*1.609344] {total_risk_SE.txt};
    \addplot+ [mark=none] table [x=x,y expr=\thisrow{SE}*1.609344] {total_risk_SE.txt};
    \addplot+ [mark=none, gray, dashed, domain=500:150000, samples=2] {0.05 * 13.226341374847571 * 1.609344};

    \addlegendentry{Variance}
    \addlegendentry{Coverage}
    \addlegendentry{Total SE}
    \addlegendentry{5\% of total risk}

\end{axis}
\end{tikzpicture}
\caption{Evolution of the SE of the total risk with the number of samples}
\label{fig:SE_total}
\end{figure}
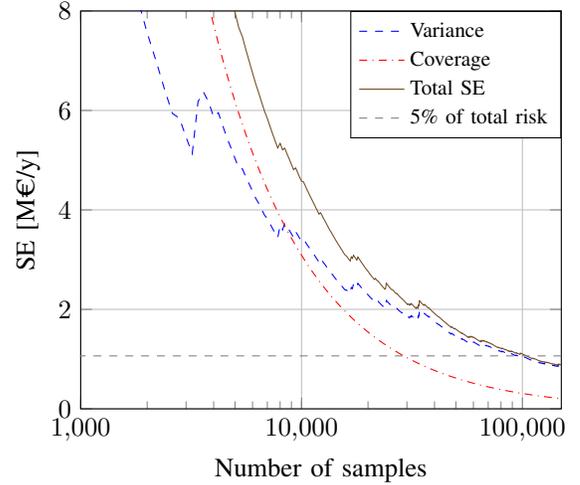

\begin{figure}
  \centering
  \begin{tikzpicture}
  \pgfplotsset{width=0.9\linewidth}
  \begin{axis}[
      height=5cm,
      xlabel=Contingency ID,
      xmin=0.5, xmax=15.5,
      ymin=0,
      yticklabel style={/pgf/number format/fixed},
      ylabel=Risk {[M€/y]},
      xtick={1,2,...,15},
  ]

  \addplot+ [ybar, ybar legend, fill, mark=none, bar width=5pt] table [x=id,y expr=\thisrow{cost}*1.609344] {critical_SE.txt};
  \addplot+ [mark=none, domain=0:21] {0.01 * 13.226341374847571 * 1.609344};

  \addlegendentry{Risk}
  \addlegendentry{\(SE_i\)}

  \end{axis}
  \end{tikzpicture}
  \caption{Risk of the 15 most critical contingencies and associated SE}
  \label{fig:SE_individual}
\end{figure}
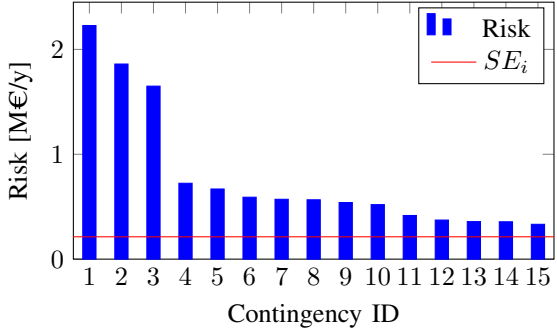

It can be noted that, for a given number of samples, the statistical accuracy of the total risk estimate is better than the one of the individual contingencies. Indeed, Figure~\ref{fig:SE_total} shows that with 150,000 samples, the SE of the total risk is smaller than 5\%. On the other hand, Figure~\ref{fig:SE_individual} shows a SE higher than 50\% for all contingencies except the 10 most critical ones. This figure shows that, with the chosen value of \(\epsilon\) (1\%), the 10 most critical contingencies are most likely correctly identified. The individual risk associated with the remaining contingencies is close to or smaller than \(SE_i\), so there is a chance that contingencies with a higher risk than those remaining contingencies have been missed in the analysis. Instead of \(SE_i < \epsilon R\), it would be possible to define the stopping criteria such that \(SE_i\) must be smaller than, \eg, the risk of the tenth most critical contingency. This would remove the need to manually set a value for \(\epsilon\), however, this could cause the computation to never or slowly converge in case the tenth most critical contingency has a very low risk contribution (\ie if the risk is dominated by the first nine contingencies).

\subsection{Screening}
\label{sec:results_screening}

We now study how the addition of a screening process impacts the accuracy and computational burden of the PDSA. The main results are given in Table~\ref{tab:screening}. First, it is important to notice that even though we consider relatively severe contingencies (compared to the N-1 contingencies with normal clearing time the system has been designed to withstand), only 3\% of the sampled scenarios (4329 out of around 152,138) lead to consequences (around 2\% for delayed-clearing N-1 scenarios, and 4\% for N-2 scenarios). Therefore, without screening, 95\% of the computation time is wasted on secure scenarios (not 97\%, as an unsecure scenario takes on average more time to simulate than a secure one).

With perfect screening, the computational burden of the PDSA could thus be reduced by a factor 20. The screening process used in this work does not reach this performance however. It has a low false negative (FN) rate, \ie very few unsecure scenarios are missed, so it has a low impact on the PDSA accuracy (only 4\% of the total risk is missed). However, it has a high false positive (FP) rate, \ie many secure scenarios are flagged as unsecure and therefore unnecessarily simulated. Screening thus only speeds up the PDSA by a factor 1.94, way less than the theoretical limit of 20.

\begin{table}
  \centering
  \caption{Performance of the screening process and impact on the PDSA accuracy and computation time}
  \label{tab:screening}
  \begin{adjustbox}{max width=\linewidth}
  \begin{tabular}{@{}lllllll@{}}
    \toprule
    \multirow{2}{*}{\begin{tabular}[c]{@{}l@{}}Contin-\\ gencies\end{tabular}} &
      \multicolumn{2}{l}{Unsecure cases} &
      \multicolumn{2}{l}{Secure cases} &
      \multirow{2}{*}{\begin{tabular}[c]{@{}l@{}}Missed\\ risk (\%)\end{tabular}} &
      \multirow{2}{*}{Speed-up} \\ \cmidrule(lr){2-3} \cmidrule(lr){4-5}
        & FN & TP   & FP     & TN     &     &      \\ \midrule
    N-1 & 4  & 1553 & 40,100 & 52,400 & 0.3 & 2.01 \\
    N-2 & 81 & 1847 & 22,300 & 25,200 & 6.4 & 1.81 \\
    All & 85 & 3400 & 62,400 & 77,800 & 4.0 & 1.94 \\ \bottomrule
  \end{tabular}
  \end{adjustbox}
\end{table}

This is mainly because the EEA underestimate CCT in this case due to the large penetration of inverter-based generators. Indeed, due to their low capacity factors, inverter-based generators often have room to provide fast voltage support which helps with the angular stability of synchronous generators. Table~\ref{tab:CCT_margin} shows that when using a CCT margin of -20ms in the screening process, \ie assuming that faults cleared in 200ms are secure if the EEA predicts a CCT larger than 180ms, a speed-up of 3.5 can be obtained but at the cost of missing 16.6\% of the total risk. Better stability indicators should be developed if higher speed-ups and/or lower impact on accuracy are desired.

\begin{table}
  \centering
  \caption{Impact of the CCT Margin on screening performance}
  \label{tab:CCT_margin}
  \begin{adjustbox}{max width=\linewidth}
  \begin{tabular}{@{}lll@{}}
  \toprule
  CCT margin (ms) & Missed risk (\%) & Speed-up \\ \midrule
  50  & 4.0  & 1.94 \\
  0   & 9.3  & 2.75 \\
  -20 & 16.6 & 3.49 \\ \bottomrule
  \end{tabular}
  \end{adjustbox}
\end{table}

\subsection{Handling of uncertain protection behaviour during fast cascading outages}
\label{sec:results_protection}

The results discussed above have been obtained by running 5 MC simulations for each scenario for which protection-related uncertainties are expected to have an impact. This was the case for a fifth (834 out of 4329) of unsecure scenarios. Of these, half (408 out of 834) led to different consequences depending on the sampled protection system parameters. In the remaining half, the cascading path was affected by protection-related uncertainties, but the final consequences were not. There are two main reasons for this. The first is that changing the order of protection operations does not always impact the general evolution of the cascade. The second is that the operation of an additional protection system does not necessarily impact the consequences, for example, if it occurs in an island of the system that will nevertheless collapse. This is discussed in more details in our previous work~\cite{HandlingProtections}.

Performing 5 MC simulations for each scenario impacted by protection-related uncertainties can be viewed as a way to perform importance sampling. However, as shown above, a crude MC approach is more efficient for most contingencies. Theoretically, it would thus be more efficient to draw one sample of protection-related parameters for each sample of operating conditions. But in practice, the proposed approach gives more intuitive results because it allows one to separate the impact of operating conditions and of protection parameters when interpreting the results.

Also, it can be argued that using an indicator to predict which scenarios are sensitive to protection-related uncertainties increases coverage as one sample directly accounts for all possible values of protection parameters, reducing the dimension of the uncertainty space and the likelihood of missing critical regions. In any case, the impact on computation time is relatively limited as the scenarios that are secure and not affected by protection-related uncertainties take most of the computation time.

\subsection{Security enhancement}
\label{sec:results_ML}

Table~\ref{tab:critical} shows the risk associated with the 10 most critical contingencies. It shows that those 10 critical contingencies (out of 708) contribute to more than 40\% of the total risk, and that particular attention should thus be given to those contingencies. As discussed in section~\ref{sec:ML}, data-mining techniques can be used to estimate the security boundary, \ie limit between secure and unsecure operating conditions, for given contingencies based on the results of the PDSA.

\begin{table}
\centering
\caption{Most critical contingencies}
\label{tab:critical}
\begin{tabular}{@{}lll@{}}
\toprule
Branch 1 & Branch 2 & Risk (M€/y) \\ \midrule
A34    & /         & 2.22 \\
A25-1  & A25-2     & 1.87 \\
CA-1   & /         & 1.66 \\
A23    & A28       & 0.72 \\
CB-1   & /         & 0.68 \\
C22    & /         & 0.60 \\
A34    & A25       & 0.56 \\
AB3    & /         & 0.56 \\
A25    & /         & 0.55 \\
A34    & CA-1      & 0.51 \\
Others & /         & 12.4 \\ \bottomrule
\end{tabular}
\end{table}

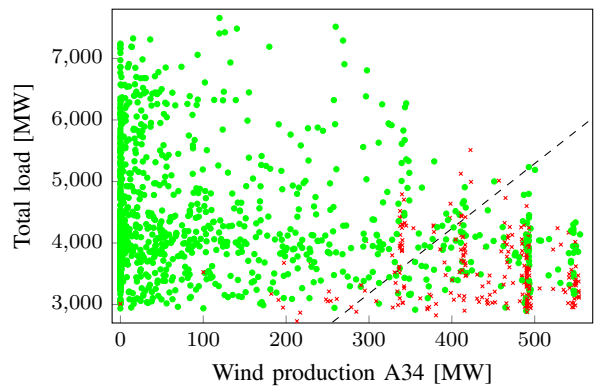
\begin{figure}
\centering
\begin{tikzpicture}
\begin{axis}[
    group style={%
        columns=2,
        group name=plots,
        xlabels at=edge bottom,
        y descriptions at=edge left,
        vertical sep = 5pt,
        horizontal sep = 5pt,
    },
    footnotesize,
    xlabel = {Wind production A34 [MW]},
    ylabel = {Total load [MW]},
    tickpos=left,
    xtick={0,100,200,300,400,500},
    xmin=-10,
    xmax=570,
    ymin=2700,
    ymax=7800,
    ytick={3000,4000,5000,6000,7000},
    width=0.9\linewidth,
    height=0.65\linewidth,
]
\addplot+[scatter, only marks,
  scatter/classes={0={mark=*,green, mark size=1pt},
                   1={mark=x,red, mark size=1pt},
                   2={mark=*,green, mark size=1pt},  
                   3={mark=x,red, mark size=1pt}  
                  },
  scatter src=explicit symbolic] table [x=wind, y=load, meta=color] {A34.txt};
\addplot+[black, dashed, no marks, domain=0:600]{(0.39547506181950903 * x + 0.29077603410661323) / 0.03742073838460856};
\end{axis}
\end{tikzpicture}
\caption{Safe (green dots) and unsafe (red crosses) operating conditions for faults on line A34 (with delayed clearing) and SVM prediction (dashed line)}
\label{fig:A34}
\end{figure}

Fig.~\ref{fig:A34} demonstrates this for the contingency of line A34 (most critical contingency), a line that connects a large wind farm located in the South of the system as shown in Fig.~\ref{fig:RTS}. The x- and y-axis of Fig.~\ref{fig:A34} (power production at the large wind farm and total system load) are the two features that have been selected by the sequential feature selection process. The dashed line is the security boundary estimated by an SVM. This figure helps to understand the results of the PDSA as it suggests that the system tends to be less secure (for this contingency) when the wind farm connected through line A34 is producing high amounts of power and when the total load is low.

Additional information can also be gathered from the sequential feature selection procedure. Indeed, Table~\ref{tab:sequential_feature} shows the most important features identified in the first iteration for the contingency of line A34. (The accuracies listed in Table~\ref{tab:sequential_feature} are the accuracy of a single-feature SVM using said feature.) The most important feature is the wind production near line A34 (88.0\% accuracy), but the power flows in lines A34 and A30 are very close (86.5 and 85.8\% accuracy respectively). This indicates that the loss of stability following the contingency of line A34 is likely caused by loss of transient stability due to high exports from the large wind plant in the area. This has indeed been checked from time-domain simulations (by manually investigating a couple of random scenarios from the PDSA results). Such information would be useful to help operators decide on some risk-reduction actions, for example in this case, installation of series capacitors, new lines, or curtailment of the wind farm (\eg via a system integrity protection scheme to only curtail following contingencies).


The total load only appears as an important feature at the second iteration of the sequential feature selection algorithm. Adding this feature to the SVM increases by 2\% (from 88.0 to 90.0\%). So the contribution of this feature is relatively minor compared to the wind production.

\begin{table}
\centering
\caption{Best features at the first iteration of the sequential feature selection procedure for the contingency of line A34 (with delayed clearing)}
\label{tab:sequential_feature}
\begin{tabular}{@{}lll@{}}
\toprule
Feature & Accuracy (\%) \\ \midrule
Wind production at bus 122 & 88.0          \\
Power flow in line A30     & 86.5          \\
Power flow in line A34     & 85.8          \\
Total wind production      & 81.4          \\ \bottomrule
\end{tabular}
\end{table}

%

\section{Applicability to large grids}
\label{sec:scalability}

In this work, we performed a PDSA on a medium-scale test grid considering 114 N-1 and 594 N-2 contingencies which took a non-negligible amount of computation time. Without screening, it takes around 400 core-hours. Such analysis can be performed in 1 days using a 16-core workstation, or a few hours in an HPC environment for an approximate cost of 120€ (assuming a renting cost of 0.3€ per core-hour).

The computation cost of the method scales primarily with the number of considered contingencies and the computation time per simulation as the stopping criterion (\ref{eq:stop}) has to be satisfied for each contingency. However, for a system of a given size, the computation time can also significantly vary depending on the value of the risk (more samples are required to assess the risk of a very safe system), the value of \(\epsilon\), and on the importance of uncertainties.

Based on table~\ref{tab:summary-N1N2}, it can be assumed that it takes around 800 simulations per delayed clearing N-1 contingency and 100 simulations per N-2 contingency to perform a complete PDSA. This assumption can be used to estimate the computational requirements for an analysis on a larger system. For example, according to~\cite{EurostagDSA}, the French power system has a little under 2000 N-1 contingencies that can simulated in less than 60s each. If we additionally assume that there are 10,000 N-2 contingencies and that they can be simulated in 120s, a PDSA would take 26,000 core-hours for N-1 contingencies (10 times less than in~\cite{EurostagDSA}) and 33,000 core-hours for N-2 contingencies, so around 18k€ in HPC. However, this costs only applies when the analysis is performed for the first time. On subsequent runs, screening techniques can be used to significantly reduce computation time (by a factor 2 in this case, but up to an order of magnitude with better screening indicators). Also, only critical contingencies identified in the first study could be rerun, reducing computation time by one or two additional orders of magnitude.

\section{Conclusion}
\label{sec:conclusion}

In this paper, we proposed a methodology for probabilistic dynamic security assessment and enhancement of power systems that considers N-k contingencies, load and generation variability, and uncertain cascade propagation caused by uncertain protection system behaviour. In this methodology, a database of likely operating conditions is generated via weather data, a market model and a model of operators' preventive actions. The database is then sampled along with the uncertainty of protection system behaviour, and dynamic simulations of (N-1 and N-k) contingencies are performed to assess the security of the system. Optionally, a screening of contingencies can be performed to limit the computational burden of the analysis. Finally, support vector machines and sequential feature selection are used to ease the interpretation of the results by identifying the root causes of the risk from the most critical contingencies.

The proposed method is applied on the RTS-GMLC system in an HPC environment. The method is able to identify critical contingencies, \ie contingencies that have a significant contribution to the total risk. The computational burden of the method is high but manageable: 400 core-hours are required to perform the analysis on the RTS-GMLC, and we estimate that around 60,000 core-hours would be required for a large system with a contingency list of 12,000 contingencies. Moreover, we show that this burden can significantly be reduced with the use of screening techniques and through proper choice of statistical accuracy requirements.

The contingencies considered in this work are line faults followed a failure to trip of some protection system. Unwanted trips are another important failure mode of protection systems that can transform an N-1 contingency into an N-k contingency. There is however very limited literature on the modelling of those unwanted trips (for modern numerical protection relays) which is something we will study in future work. Also, better stability indicators could be developed for systems with high shares of inverter-based generation and for severe contingencies to improve the screening process and reduce the computational burden of the PDSA.

\section*{Acknowledgments}
Computational resources have been provided by the Consortium des Équipements de Calcul Intensif (CÉCI), funded by the Fonds de la Recherche Scientifique de Belgique (F.R.S.-FNRS) under Grant No. 2.5020.11 and by the Walloon Region.

\bibliographystyle{IEEEtran}
\bibliography{IEEEabrv,bib}

\vfill

\end{document}